\documentclass[fleqn,12pt,twoside]{article}
\usepackage{espcrc1}

\usepackage{graphicx}

\def\bbox#1{\hbox{\boldmath$#1$}}

\title{The Role of Produced Hadrons in $J/\psi$ Suppression\thanks
{This research was supported by the Division of Nuclear Physics,
Department of Energy, under Contract No.\ DE-AC05-00OR22725 managed by
UT-Battelle, LLC.  ES acknowledges support from the DOE under grant
DE-FG02-00ER41135 and DOE contract DE-AC05-84ER40150 under which the
Southeastern Universities Research Association operates the Thomas
Jefferson National Accelerator Facility.}
}

\author{Cheuk-Yin Wong\address{Physics Division, Oak Ridge National
Laboratory, Oak Ridge, Tennessee 37831 USA\\ 
$^{\rm b}$Department of Physics, University of Tennessee, Knoxville,
TN 37996 USA\\
$^{\rm c}$Department of Physics and Astronomy, Univ. of Pittsburgh,
Pittsburgh, PA 15260 USA\\
$^{\rm d}$Jefferson Lab, Newport News, VA 23606 USA\\
$^{\rm e}$Department of Physics, Univ.  of Tennessee Space
Institute, Tullahoma, TN 37388 USA }, 
Ted Barnes$^{\rm a,b}$, 
Eric S. Swanson$^{\rm c,d}$, 
and Horace W. Crater$^{\rm e}$ }

\begin{document}

\maketitle

\vspace*{-0.1cm}

\begin{abstract}

{ Hadrons produced in high-energy heavy-ion collisions can suppress
the production of $J/\psi$ and other charmonium states by charmonium
spontaneous dissociation as the hot hadronic environment alters the
interaction between the charm quark and charm antiquark.  Furthermore,
hadrons can thermalize a charmonium to excite it to higher charmonium
states which subsequently dissociate spontaneously.  They can also
collide with a charmonium to lead to its prompt dissociation into an
open charm pair.}
\end{abstract}

\vspace{-0.3cm}

\section{Introduction}

The suppression of heavy quarkonium production has been the subject of
intense studies as it was proposed as a signature for the quark-gluon
plasma \cite{Mat86}.  Recent experimental observations of an anomalous
$J/\psi$ suppression in Pb+Pb collisions by the NA50 Collaboration
\cite{Gon96} have aroused a great deal of interest \cite{Won96}.
There is, however, considerable uncertainty concerning the role of the
produced hadrons on the suppression of $J/\psi$ production.  In order
to confirm that the suppression of $J/\psi$ comes from the presence of
the quark-gluon plasma, it is necessary to understand how the produced
hadrons may affect the suppression of $J/\psi$ production so that the
desired signal from the quark-gluon plasma can be separated out.

We shall study three different mechanisms of $J/\psi$ and charmonium
dissociation in a hadronic medium: spontaneous dissociation,
dissociation by thermalization, and dissociation by collision.
We shall briefly summarize these mechanisms in turn.

\vspace{-0.3cm}
\section{ Interaction between the Charm Quark and Charm Antiquark }
\vspace*{-0.1cm}

The interaction between the quark and the antiquark in a heavy
quarkonium depends on the temperature of the medium \cite{Kar00}.  We
can understand such a dependence as coming from the change of the
quark and gluon fields between the heavy quark pair and the change of
the QCD vacuum surrounding the quarkonium.  The former arises from the
disorientation of the quark and gauge fields between the heavy-quark
pair as temperature increases, while the latter arises from the change
of the external pressure which tends to confine the quark with the
antiquark.  At the deconfinement phase transition temperature, the
confining interaction is expected to vanish.  Hadron matter at high
temperature provides an altered environment which can lead to the
spontaneous dissociation of a heavy quarkonium.

Using the $Q$-$\bar Q$ interaction as inferred from lattice gauge
calculations of Karsch $et~al.$\cite{Kar00}, Digal, Petreczky, and
Satz \cite{Dig01a} recently reported theoretical results on the
dissociation temperatures of heavy quarkonia. Subsequent analysis
using different parametrization of the interaction and selection rules
was given in \cite{Won01c} and will be described below.

To study the behavior of a heavy quarkonium at finite temperatures, we
calculate the energy $\epsilon$ of the heavy quarkonium state $(Q\bar
Q)_{JLS}$ from the Schr\" odinger equation \cite{Won01c,Won99}
\begin{eqnarray}
\label{eq:sch}
\biggl \{\! - \nabla \cdot { \hbar^2 \over 2\mu_{12}}\nabla   
+ V(r,T)
+(m_Q+m_{\bar Q}-M_{(Q\bar q)}-M_{(q \bar Q)})\theta (R-r)
 \!\biggr \}
\psi_{JLS} (\bbox{r})
= \epsilon \psi_{JLS} (\bbox{r}),
\end{eqnarray}
where the energy $\epsilon$ is measured relative to the pair of
lowest-mass mesons at $r\to \infty$, $M_{(Q\bar q)}$ and $M_{(q \bar
Q)}$ are the masses of the open heavy-quark mesons, and  $R \sim 0.8$ fm.
We represent the interaction $V(r,T)$, as inferred from lattice gauge
calculations of Karsch $et~al.$\cite{Kar00}, by a Yukawa plus an
exponential potential \cite{Kar88,Won99}
\begin{eqnarray}
\label{eq:pot}
V(r,T)=-{4 \over 3} {\alpha_s e^{-\mu (T) r}\over r}
    -{ b(T) \over \mu(T)} e^{-\mu(T) r}
~~~~~~~~~{\rm (2)}
\nonumber
\end{eqnarray}

\hangafter=-6
\hangindent=-2.8in
\noindent
where $b(T)=b_0[1-(T/T_c)^2]\theta(T_c-T)$, $b_0=0.35$ GeV$^2$,
$\mu(T)=0.28$ GeV, and $T_c$ is the deconfinement phase transition
temperature.  We use a running $\alpha_s$ and include spin-spin,
spin-orbit, and tensor interactions \cite{Won01c}.  The eigenvalues of
the Hamiltonian can be obtained by matrix diagonalization.

\vspace*{-1.1cm}
\hspace*{3.3in}
\includegraphics{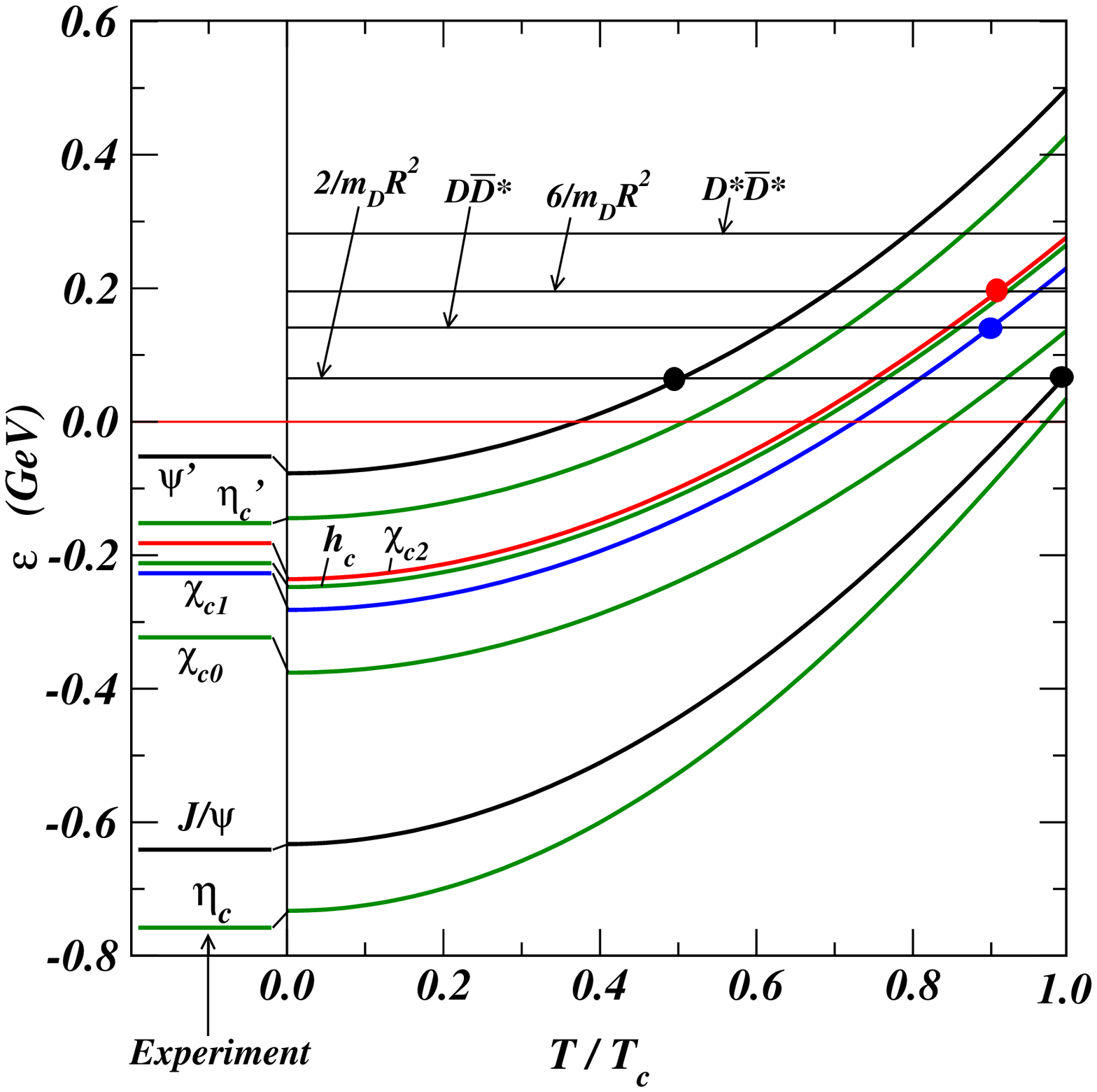}

\vspace*{4.2cm}
\hspace*{3.6in}
\begin{minipage}[t]{7 cm}
\noindent {\bf Fig.\ 1}.  {Charmonium states as a
function of temperature. }
\end{minipage}
\vskip 2truemm
\noindent 

\vspace*{-4.5cm}
\section{Spontaneous Dissociation}
\vspace*{-0.1cm}

\hangafter=-16 \hangindent=-2.8in Figure 1 shows the charmonium state
energy as a function of temperature~\cite{Won01c}.  The solid circles
indicate the locations of the dissociation temperatures after taking
into account the selection rules.  The dissociation temperatures are
listed in Table I.

\vspace*{0.8cm}
\hspace*{3.2in}
\begin{minipage}[t]{7cm}
\hspace*{0.1cm}\noindent
{\bf Table I.}  {Charmonium  dissociation temperatures $T_d$ in
units of $T_c$}
\end{minipage}

\vspace*{0.1cm}
\hspace*{2.9in}
\begin{tabular}{|c|c|c|c|c|}\hline
{\rm Charmonium}& $~\psi'~$ &$~\chi_{c2}~$ 
                      & $~\chi_{c1}~$ & $~J/\psi~$ 
                        \\
\hline
 $T_d/T_c$             & 0.50 & 0.91 & 0.90&  0.99 \\
\hline  
\vspace*{-0.0cm}
 $T_d/T_c$         & 0.1-0.2 & 0.74 & 0.74& 1.10 \\
(Digal $et~al.$)     &  &  &  & \\
\hline
\end{tabular}

\vspace*{-3.7cm} \hangafter=-16 \hangindent=-3.4in We confirm the
general features of the results of Digal $et~al.$\cite{Dig01a}, but
there are also differences as the dissociation temperatures depend on
the selection rules and the spin-orbit, spin-spin, and other details
of the potential.  Our results from Fig. 1 and Table I indicate that
the dissociation temperatures of all charmonia, obtained by using the
temperature-dependent potential of Eq.\ (\ref{eq:pot}), are below
$T_c$.

\newpage

\section{Dissociation by Thermalization}
\vspace*{-0.1cm}

\hangafter=-11 \hangindent=-2.9in If a quarkonium is placed in a hadronic
medium, there will be non-dissociative reactions between the
quarkonium and medium particles which\break change a charmonium state
into another charmonium state: $h + (Q\bar Q)_{JLS} \leftrightarrow h'
+ (Q\bar Q)_{J'L'S'}$.  These reactions lead to the thermalization of
the charmonium system \cite{Kha95,Won01c}.  When the heavy quarkonium
system is in thermal equilibrium with the medium, the occupation
probabilities of the heavy quarkonium state $\epsilon_i$ will be
distributed according to the Bose-Einstein distribution

\vspace*{1.2cm}
\hspace*{3.2in}
\includegraphics{evap.S.si.eps}

\vspace*{-2.0cm}
\hspace*{3.5in}
\begin{minipage}[t]{6.8cm}
\noindent {\bf Fig.\ 2}.  {The survival probability of charmonium as a
function of $T/T_c$.}
\end{minipage}
\vskip 4truemm
\noindent 

\vspace*{-1.8cm}
\noindent
\begin{eqnarray}
n_i={1\over \exp\{(\epsilon_i-\mu)/T\}-1},~~~~~~~~~~~~~~~~~~~~~~~{\rm (3)}
\nonumber
\end{eqnarray}

\vspace*{-0.3cm}
\noindent
where $\mu$ is the chemical potential.  The system at thermal
equilibrium will lose the memory of the initial state.  There is the
probability $f$ for the quarkonium to lie above the threshold for
spontaneous dissociation leading to its subsequent dissociation into
an open charm pair.  We evaluate the survival probability $S=1-f$ as a
function of temperature for charmonium.  The results are shown in
Fig.\ 2.  A state label along the curve denotes the onset of the
occurrence when that state emerges above its dissociation threshold.
As one observes, the survival probability $S$ decreases with
increasing temperature in a step-wise manner, and it becomes
quite small as one approaches the transition temperature.

\vspace*{-0.5cm}
\section{Dissociation by Collision with Hadrons}
A heavy quarkonium can dissociate into an open charm pair by collision
with hadrons through reactions of the type $(q\bar q) + (Q\bar
Q)_{JLS} \to (Q \bar q) + (q \bar Q)$.  As the temperature of the
medium increases, the quarkonium state energy changes (Fig. 1) and the
dissociation threshold energy decreases.  As a consequence, the
dissociation cross section will increase.  We calculate the
dissociation cross sections of $J/\psi$ in collision with $\pi$ as a
function of the temperature using the Barnes and Swanson model
\cite{Bar92}, as in Ref.\ ~\cite{Won01}.  We use the
temperature-dependent Yukawa and exponential interaction in Eq.\
(\ref{eq:pot}) with the usual color dependence for the interquark
interaction\cite{Won01c}.  The sum of dissociation cross sections for
$\pi+J/\psi \to D\bar D^*, D^*\bar D, D^* \bar D^* $ are shown in
Fig. 3 for different temperatures $T/T_c$ as a function of the
kinetic energy $E_{KE}$.

We observe in Fig.\ 3 that the maximum values of the dissociation
cross sections increase and the positions of the maxima shift to lower
kinetic energies as the temperature increases. Such an increase arises
from the decrease of the threshold energies as the temperature
increases.  Over a large range of temperatures below the phase
transition temperature, dissociation cross sections of $J/\psi$ in
collisions with $\pi$ are large.

We can estimate the survival probability of a heavy quarkonium in a
hot pion gas in the presence of this type of collisional dissociation.
The degree of absorption by collision 
with pions depends on the
initial absorption time $\tau_0$, the freeze-out pion density
$\rho_{\rm freeze}$, and the initial temperature $T_0$.  If $T_0 \sim
T_c$, $\rho_{\rm freeze}=0.5/$ fm$^3$, and $\tau_0=3$ fm/c$+R/\gamma$,\break
then for the most central Pb-Pb collision at 158A GeV, the heavy
quarkonium survival 

\hspace*{3.0in}
\includegraphics{pi.si.T.run.yuex.eps}

\vspace*{3.4cm}
\hspace*{3.2in}
\begin{minipage}[t]{7cm}
\noindent {\bf Fig.\ 3}.  {Total dissociation cross section of
$\pi+J/\psi$ for various temperatures as a function of the kinetic
energy $E_{KE}$. }
\end{minipage}
\vskip 2truemm
\noindent 

\vspace*{-5.9cm}
\hangafter=-13 \hangindent=-3.1in 
\noindent
probability is $S\sim 0.5$, and for the most central Au-Au collision
RHIC at $\sqrt{s_{NN}}=200$ GeV, the heavy quarkonium survival
probability is $S\sim 0.1$.  These estimates show that the absorption
of $J/\psi$ by the hot pion gas is substantial.

\vspace{-0.3cm}
\section{Discussions and Conclusions}
\vspace*{-0.1cm}

\hangafter=-4 \hangindent=-3.1in 
The temperature of the medium alters the interaction between a heavy
quark and antiquark in the medium.  We have calculated
the quarkonium state energies using a potential inferred from lattice
gauge calculations of Karsch $et~al.$\cite{Kar00} and obtain the
dissociation temperatures.  We find that the dissociation temperatures
of all charmonia are below $T_c$.

A quarkonium in a medium can collide with particles in the medium to
reach thermal equilibrium.  A heavy quarkonium in thermal equilibrium
can dissociate by thermalization as there is a finite probability for
the system to be in an excited state lying above its dissociation
threshold.  We find that the survival probability of a charmonium
decreases with increasing temperatures in a step-wise manner.

Dissociation of a heavy quarkonium into open heavy-quark mesons can
occur in collision with hadrons.  As the temperature increases, the
threshold energies for collisional dissociation decrease.  As a
consequence, the dissociation cross sections increase.  We have
estimated the absorption of $J/\psi$ in collision with pions in
central Pb-Pb collisions at SPS and RHIC energies and found the
absorption to be substantial.  Further microscopic investigations of
the dissociation of heavy quarkonium in collision with hadrons in
high-energy heavy-ion collisions will be of great interest.

\end{document}